\documentclass{article}
\usepackage[utf8]{inputenc}
\usepackage[T1]{fontenc}
\usepackage{amsmath}
\usepackage{amsfonts}
\usepackage{latexsym}
\usepackage{amssymb}
\usepackage{times}
\usepackage{ifpdf}
\usepackage{makeidx}
\usepackage{graphicx}
\ifpdf
 \usepackage[pdftex,colorlinks]{hyperref}
\else
 \usepackage[ps2pdf,breaklinks=true,colorlinks=true,linkcolor=red,citecolor=green]{hyperref}
 \fi

\newcommand{\Z}{{\mathbb{Z}}}

\newcommand{\Q}{{\mathbb{Q}}}

\newtheorem{thm}{Theorem}

\title{A probabilistic and deterministic modular algorithm for computing Groebner basis over $\Q$.}

\author{Bernard Parisse\\Institut Fourier\\UMR 5582 du
  CNRS\\Universit\'e de Grenoble I}
\date{2013}

\begin{document}

\maketitle

\begin{abstract}
Modular algorithm are widely used in computer algebra systems (CAS), for
example to compute efficiently the gcd of multivariate polynomials. It
is known to work to compute Groebner basis over $\Q$, but it does not seem to be
popular among CAS implementers. In this paper, I will show how
to check a candidate Groebner basis (obtained by
reconstruction of several Groebner basis modulo distinct prime numbers)
with a given error probability, that may be 0 if a certified
Groebner basis is desired.
This algorithm is now the default algorithm used by
the Giac/Xcas computer algebra system with competitive timings,
thanks to a trick that can accelerate computing
Groebner basis modulo a prime once the computation has been done
modulo another prime.
\end{abstract}

\section{Introduction}
During the last decades, considerable improvements have been made 
in CAS like Maple or specialized systems like Magma, Singular, 
Cocoa, Macaulay... to compute Groebner basis. 
They were driven by
implementations of new algorithms speeding up the original Buchberger (\cite{buchberger1985grobner})
algorithm: Gebauer and M\"oller criterion (\cite{Gebauer1988275}), F4 and F5
algorithms from J.-C. Faug\`ere (\cite{F99a}, \cite{Fau02a}), and are widely described in the
literature if the base field is a finite field.
Much less was said about computing over $\Q$. It seems that
implementers are using the same algorithm as for finite fields,
this time working with coefficients in $\Q$ or in $\Z$ (sometimes
with fast integer linear algebra), 
despite the fact that an efficient p-adic or Chinese remaindering
algorithm were described as soon as in year 2000 by E. Arnold
(\cite{Arnold2003403}). 
The reason might well be that these modular algorithms suffer from a
time-consuming step at the end: checking that the reconstructed
Groebner basis is indeed the correct Groebner basis.

Section 2
will show that if one accepts a small error probability, this check 
may be fast, so we can let the user choose between a fast conjectural
Groebner basis to make his own conjectures and a slower certified
Groebner basis once he needs a mathematical proof.

Section 3 will explain learning, a process that can accelerate the
computation of a Groebner basis modulo a prime $p_k$ once the
same computation but modulo another prime $p$ has already been
done ; learning is an alternative to the F5 algorithm
in order to avoid computing useless critical pairs that reduce to 0. The
idea is similar to {\tt F4remake} by Joux-Vitse (\cite{joux2011variant}) 
used in the context of computing Groebner basis in large finite fields.

Section 4 will show in more details how the gbasis algorithm is implemented
in Giac/Xcas (\cite{giac}) and show that - at least for the classical academic
benchmarks Cyclic and Katsura - the deterministic modular algorithm
is competitive or faster than the best open-source implementations
and the modular probabilistic algorithm is comparable to Maple
and slower than Magma on one processor (at least for moderate integer coefficient
size) and may be faster than Magma on multi-processors, 
while computation modulo $p$ are faster for characteristics in the
24-31 bits range.
Moreover the modular algorithm memory usage is essentially
twice the memory required to store the basis on $\Q$, sometimes
much less than the memory required by other algorithms.

\section{Checking a reconstructed Groebner basis}
Let $f_1,..,f_m$ be polynomials in $\Q[x_1,..,x_n]$, $I=<f_1,...,f_m>$
be the ideal generated by $f_1,...,f_n$. Without loss of generality, we may
assume that the $f_i$ have coefficients in $\Z$ by multiplying
by the least common multiple of the denominators of the coefficients
of $f_i$. We may also assume that the $f_i$ are primitive by dividing
by their content.

Let $<$ be a total monomial ordering (for example {\tt revlex} the
total degree reverse lexicographic ordering). We want to compute
the Groebner basis $G$ of $I$ over $\Q$ (and more precisely
the inter-reduced Groebner basis, sorted with respect to $<$).
Now consider the ideal $I_p$ generated by the same $f_i$ but with
coefficients in $\Z/p\Z$ for a prime $p$. Let $G_p$ be the Groebner basis of $I_p$
(also assumed to be inter-reduced, sorted with respect to $<$, and with
all leading coefficients equal to 1).

Assume we compute $G$ by the Buchberger
algorithm with Gebauer and M\"oller criterion, and we reduce in $\Z$
(by multiplying the s-poly to be reduced by appropriate leading
coefficients), if no leading coefficient in the polynomials are
divisible by $p$, we will get by the same process but computing modulo
$p$ the $G_p$ Groebner basis. Therefore the computation can be
done in parallel in $\Z$ and in $\Z/p\Z$ except for a finite 
set of {\em unlucky} primes (since the number of intermediate polynomials
generated in the algorithm is finite). If we are choosing our primes
sufficiently large (e.g. about $2^{31}$), the probability to fall on
an unlucky prime is very small (less than the number of generated
polynomials divided by about $2^{31}$, even for really large 
examples like Cyclic9 where there are a few $10^4$ polynomials 
involved, it would be about {\tt 1e-5}).

The Chinese remaindering algorithm is as follow: compute $G_p$ for
several primes, for all primes that have the same leading monomials
in $G_p$ (i.e. if coefficient values are ignored), reconstruct
$G_{\prod p_j}$ by Chinese remaindering, then reconstruct a
candidate Groebner basis $G_c$ in $\Q$ by Farey reconstruction. Once it
stabilizes, do the checking step described below, and return $G_c$
on success.

{\bf Checking step}~: check that the original $f_i$ polynomials reduce
to 0 with respect to $G_c$ and check that $G_c$ is a Groebner basis.

\begin{thm} (Arnold)
If the checking step succeeds, then $G_c$ is the Groebner basis of $I$.
\end{thm}

This is a consequence of ideal inclusions (first check) and dimensions (second
check), for a complete proof, see \cite{Arnold2003403}.

{\bf Probabilistic checking algorithm}: instead of checking that s-polys
of critical
pairs of $G_c$ reduce to 0, check that the s-polys reduce to 0 
modulo several primes that do not divide the leading coefficients of
$G_c$ and stop
as soon as the inverse of the product of these primes is less than a
fixed $\varepsilon>0$.

{\bf Deterministic checking algorithm}: check that all s-polys
reduce to 0 over $\Q$. This can be done either by integer computations
(or even by rational computations, I have not tried that),
or by reconstruction of the quotients using modular reduction to 0
over $\Z/p\Z$ for sufficiently many primes. Once the reconstructed
quotients stabilize, we can check the 0-reduction identity, and this
can be done without computing the products quotients by elements of $G_c$
if we have enough primes (with appropriate bounds on
the coefficients of $G_c$ and the lcm of the denominators of the
reconstructed quotients).

\section{Speeding up by learning from previous primes}
Once we have computed a Groebner basis modulo an initial prime $p$, if $p$ is
not an unlucky prime, then we can speedup computing Groebner basis
modulo other lucky primes. Indeed, if one s-poly reduce to 0 modulo
$p$, then it reduces most certainly to 0 on $\Q$ (non zero s-poly have
in general several terms, cancellation of one term mod $p$ has
probability $1/p$, simultaneous cancellation of several terms of a non-zero
s-poly modulo $p$ is highly improbable), and we discard this s-poly in the
next primes computations. We name this speedup process {\em learning}. 
It can also
be applied on other parts of the Groebner basis computation, like the
symbolic preprocessing of the F4 algorithm, where we can reuse the
same collection of monomials that were used for the first prime $p$ 
to build matrices for next primes (see Buchberger Algorithm with F4 
linear algebra in the next section).

If we use learning, we have no certification that the computation ends up
with a Groebner basis modulo the new primes. But this is not a problem,
since it is not required by the checking correctness proof, the only
requirement is that the new generated ideal is contained in the
initial ideal modulo all primes (which is still true) and 
that the reconstructed $G_c$ is a Groebner basis.

\section{Giac/Xcas implementation and experimentation}
We describe here briefly some details of the Giac/Xcas gbasis implementation
and give a few benchmarks.

The optimized algorithm runs with revlex as $<$ ordering
if the polynomials have at most 15
variables (it's easy to modify for more variables, adding multiples of
4, but this will increase a little memory required and
slow down a little).
Partial and total degrees are coded as 16 bits integers (hence the 15
variables limit, since 1 slot of 16 bits is kept for total degree). 
Modular coefficients are coded as 31 bit integers (or 24).

The Buchberger algorithm with linear algebra 
from the F4 algorithm is implemented modulo primes smaller than $2^{31}$
using total degree as selection criterion for critical pairs.\\
{\bf Buchberger algorithm with F4 linear algebra modulo a prime}
\begin{enumerate}
\item Initialize the basis to the empty list, and a list of critical
  pairs to empty
\item Add one by one all the $f_i$ to the basis and update
the list of critical pairs with Gebauer and M\"oller criterion, 
by calling the gbasis update procedure (described below step 9)
\item Begin of a new iteration:\\
All pairs of minimal total degree are
collected to be reduced simultaneously, they are removed from
the list of critical pairs.
\item The symbolic preprocessing step begins by creating 
a list of monomials, gluing
together all monomials of the corresponding s-polys (this
is done with a heap data structure). 
\item The list of monomials is ``reduced'' by division with respect 
to the current basis,
using heap division (like Monagan-Pearce \cite{monagan2011sparse}) 
without taking care
of the real value of coefficients. This gives a list of all possible remainder
monomials and a list of all possible quotient monomials and a list
of all quotient times corresponding basis element monomial products.
This last list together with the remainder monomial list is the
list of all possible monomials that may be generated reducing
the list of critical pairs of maximal total degree, 
it is ordered with respect to $<$. We
record these lists for further primes during the first prime computation.
\item The list
of quotient monomials is multiplied by the corresponding elements of the current
basis, this time doing the coefficient arithmetic.
The result is recorded in a sparse matrix, each row has a pointer
to a list of coefficients (the list of coefficients 
is in general shared by many rows, the rows have the
same reductor with a different monomial shift), 
and a list of monomial indices (where the index 
is relative to the ordered list of possible monomials). We sort
the matrix by decreasing order of leading monomial.
\item
Each s-polynomial is written as a dense vector with respect to the
list of all possible monomials, and reduced with respect to the
sparse matrix, by decreasing order with respect to $<$.
(To avoid reducing modulo $p$ each time, we are using a dense
vector of 128 bits integers on 64 bits architectures, 
and we reduce mod $p$ only at the end of the reduction. If
we work on 24 bit signed integers, we can use a dense vector 
of 63 bits signed integer and reduce the vector if the number
of rows is greater than $2^{15}$).
\item Then inter-reduction happens on all the dense vectors representing
the reduced s-polynomials, this is dense row
reduction to echelon form (0 columns are removed first). 
Care must be taken at this step
to keep row ordering when learning is active.
\item gbasis update procedure\\
Each non zero row will bring a new entry in the current
basis (we record zero reducing pairs during the first prime iteration,
this information will be used during later iterations with other
primes to avoid computing and reducing
useless critical pairs). 
New critical pairs are created with this new entry (discarding useless
pairs by applying Gebauer-M\"oller criterion).
An old entry in the basis may be removed if it's leading monomial
has all partial degrees greater or equal to the leading monomial
corresponding degree of the new entry.
Old entries may also be reduced with respect to the new entries 
at this step or at the end of the main loop.
\item If there are new critical pairs remaining start a new iteration
  (step 3). Otherwise the current basis is the Groebner basis.
\end{enumerate}

{\bf Modular algorithm}
\begin{enumerate}
\item Set a list of reconstructed basis to empty.
\item Learning prime: Take a prime number of 31 bits 
or 29 bits for pseudo division, run the Buchberger algorithm modulo this
prime recording symbolic preprocessing data and the list of critical pairs
reducing to 0.
\item Loop begin:
Take a prime of 29 bits size or a list of $n$ primes if $n$ processors
are available. Run the Buchberger algorithm.
Check if the output has the same leading terms than one of the
chinese remainder reconstructed outputs from previous primes,
if so combine them by Chinese remaindering and go to step 4, otherwise add
a new entry in the list of reconstructed basis and continue with
next prime at step 3 (clearing all learning data is probably a good
idea here).
\item If the Farey $\Q$-reconstructed basis is not 
identical to the previous one, go to the loop iteration step 3
(a fast way to check that is to reconstruct with all primes
but the last one, and check the value modulo the last prime).
If they are identical, run the final check : the initial polynomials $f_i$ must reduce
to 0 modulo the reconstructed basis and
the reconstructed basis s-polys must reduce to 0 (this is
done on $\Q$ either directly or by modular reconstruction
for the deterministic algorithm, or checked modulo several primes
for the probabilistic algorithm). On success output the $\Q$
Groebner basis, otherwise continue with next prime at step 3.
\end{enumerate}

{\bf Benchmarks}\\
Comparison of giac (1.1.0-26) with Singular 3.1 (from sage 5.10) 
on Mac OS X.6, Dual Core i5 2.3Ghz, RAM 2*2Go:
\begin{itemize}
\item Mod timings were computed modulo \verb|nextprime(2^24)|
and modulo 1073741827 (\verb|nexprime(2^30)|).
\item Probabilistic check on $\Q$ depends linearly on log of precision, two
timings are reported, one with error probability less than \verb|1e-7|, and
the second one for \verb|1e-16|.
\item Check on $\Q$ in giac can be done with integer or modular computations
hence two times are reported.
\item \verb|>>| means timeout (3/4h or more) or memory exhausted
(Katsura12 modular \verb|1e-16| check with giac) or test not done because
it would obviously timeout (e.g. Cyclic8 or 9 on $\Q$ with Singular)
\end{itemize}
\begin{tabular}{|c|c|c|c||c|c|c|} \hline
        & giac mod $p$ & giac & singular & giac $\Q$ prob. & giac $\Q$  & singular \\ 
        & 24, 31 bits & run2 & mod $p$ & \verb|1e-7|, \verb|1e-16| & certified & $\Q$ \\ \hline
Cyclic7 & 0.5, 0.58 & 0.1&2.0 & 3.5, 4.2 & 21, 29.3 & >2700 \\
Cyclic8 & 7.2, 8.9 & 1.8 &52.5 & 103, 106 & 258, 679 & >> \\
Cyclic9 & 633, 1340 & 200 & ? & 1 day & >> & >> \\ \hline
Kat8 & 0.063, 0.074 & 0.009& 0.2 & 0.33, 0.53 & 6.55, 4.35 & 4.9\\
Kat9 & 0.29, 0.39 & 0.05 &1.37 & 2.1, 3.2 & 54, 36& 41\\
Kat10 & 1.53, 2.27 & 0.3& 11.65 & 14, 20.7 & 441, 335 &  480 \\
Kat11 & 10.4, 13.8 & 2.8 & 86.8 & 170, 210& 4610 & ? \\
Kat12 & 76, 103 & 27 & 885 & 1950, RAM & RAM & >> \\ \hline
alea6 & 0.83, 1.08& .26 & 4.18 & 202, 204& 738, >> & >1h\\
\hline 
\end{tabular}\\
This leads to the following observations~:
\begin{itemize}
\item Computation modulo $p$ for 24 to 31 bits is faster that Singular, but seems also
  faster than magma (and maple). For smaller primes, magma is 2 to 3
  times faster.
\item The probabilistic algorithm on $\Q$ is much faster than Singular on these examples.
Compared to maple16, it is reported to be faster for Katsura10, 
and as fast for Cyclic8. Compared to magma, it is about 3 to 4
times slower. 
\item If \cite{magma} is up to date (except about giac), giac is the third software and first
  open-source software to solve Cyclic9 on $\Q$. It requires 378
  primes of size 29 bits, takes a little more than 1 day, requires 5Gb
  of memory on 1 processor, while with 6 processors it takes
8h30 (requires 16Gb). The answer has integer coefficients of about 1600 digits
(and not 800 unlike in J.-C. Faug\`ere F4 article), for a little
more than 1 milliion monomials, that's about 1.4Gb of RAM.
\item The deterministic modular algorithm is much faster than Singular for Cyclic examples,
and as fast for Katsura examples. 
\item For the random last example, the speed is comparable between
  magma and giac. This is where there are less pairs reducing to
  0 (learning is not as efficient as for Cyclic or Katsura) and larger
  coefficients. This would suggest that advanced algorithms 
like f4/f5/etc. are probably
not much more efficient than Buchberger algorithm for these kind
of inputs without symmetries.
\item Certification is the most time-consuming part of the process (except
for Cyclic8). Integer certification is significantly faster than modular certification
for Cyclic examples, and almost as fast for Katsura.
\end{itemize}

Example of Giac/Xcas code:
\begin{verbatim}
alea6 := [5*x^2*t+37*y*t*u+32*y*t*v+21*t*v+55*u*v,
39*x*y*v+23*y^2*u+57*y*z*u+56*y*u^2+10*z^2+52*t*u*v,
33*x^2*t+51*x^2+42*x*t*v+51*y^2*u+32*y*t^2+v^3,
44*x*t^2+42*y*t+47*y*u^2+12*z*t+2*z*u*v+43*t*u^2,
49*x^2*z+11*x*y*z+39*x*t*u+44*x*t*u+54*x*t+45*y^2*u,
48*x*z*t+2*z^2*t+59*z^2*v+17*z+36*t^3+45*u];
l:=[x,y,z,t,u,v];
p1:=prevprime(2^24); p2:=prevprime(2^29);
time(G1:=gbasis(alea6 % p1,l,revlex));
time(G2:=gbasis(alea6 % p2,l,revlex));
threads:=2; // set the number of threads you want to use
// debug_infolevel(1); // uncomment to show intermediate steps
proba_epsilon:=1e-7; // probabilistic algorithm.
time(H0:=gbasis(alea6,indets(cyclic5),revlex));
proba_epsilon:=0; // deterministic
time(H1:=gbasis(alea6,indets(cyclic5),revlex));
time(H2:=gbasis(alea6,indets(cyclic5),revlex,modular_check));
size(G1),size(G2),size(H0),size(H1),size(H2);
write("Halea6",H0);
\end{verbatim}
Note that for small examples (like Cyclic5), the system performs always the deterministic
check (this is the case if the number of elements of the reconstructed basis
to 50).

\section{Conclusion}
I have described some enhancements to a modular algorithm
to compute Groebner basis over $\Q$ which, combined to 
linear algebra from F4, gives
a sometimes much faster open-source implementation 
than state-of-the-art open-source implementations 
for the deterministic algorithm. 
The probabilistic algorithm is also not ridiculous
compared to the best publicly available closed-source implementations,
while being much easier to implement
(about 10K lines of code, while Fgb is said to be 200K lines of code,
no need to have highly optimized sparse linear algebra).

This should speed up conjectures with the probabilistic algorithm
and automated proofs using the deterministic
algorithm (e.g. for the Geogebra theorem prover
\cite{botanaimplementing}), 
either using Giac/Xcas (or one of it's interfaces
to java and python) or adapting it's implementation
to other open-source systems.
With fast closed-source implementations (like maple or magma), 
there is no certification that the result is a Groebner basis :
there might be some hidden probabilistic
step somewhere, in integer linear system reduction for example. I have
no indication that it's the case but one can never know if the code is
not public, and at least for my implementation, certification
might take a lot more time than computation. 

There is still room for additions and improvements
\begin{itemize}
\item the checking step can certainly be improved using
knowledge on how the basis element modulo $p$ where
built.
\item checking could also benefit from parallelization.
\item As an alternative to the modular algorithm,
a first learning run could be done modulo a 24 bits prime, and
the collected info used for f4 on $\Q$ as a probabilistic alternative
to F5.
\item FGLM conversion is still not optimized and therefore
slow in Giac/Xcas, 
\end{itemize}


{\bf Acknowledgements} \\
Thanks to Fr\'ed\'eric Han for interfacing giac with Python.
Thanks to Vanessa Vitse for insightfull discussions.

\bibliography{gb.bib}

\end{document}